\documentclass[aps,prd,superscriptaddress]{revtex4-2}

\usepackage{enumerate}

\usepackage{amsmath}
\usepackage{amssymb}
\usepackage{amsfonts}
\usepackage{amscd}
\usepackage{mathrsfs}

\usepackage{latexsym}
\usepackage{bm}

\usepackage{graphicx}
\usepackage{epsfig}
\usepackage{extarrows}
\usepackage{hhline,multirow,array}
\usepackage{dcolumn}
\usepackage{url}
\usepackage[colorlinks=true,linkcolor=blue]{hyperref}
\usepackage{color}
\usepackage{indentfirst}
\usepackage{subfigure}
\usepackage{psfrag}
\usepackage{slashed}
\usepackage{float}
\usepackage{setspace}
\usepackage{makecell,rotating}
\allowdisplaybreaks[2]

\newcommand\lr[1]{\left(#1\right)}
\DeclareMathOperator{\tr}{Tr}

\begin{document}
\title{Nonlocal QED and lepton g-2 anomalies}
\author{Hang Li}
\affiliation{Institute of High Energy Physics, Chinese Academy of Sciences,
	Beijing 100049, China}
\affiliation{{College of Physics Sciences, University of Chinese Academy of Sciences, Beijing 100049, China}}

\author{P. Wang}
\affiliation{Institute of High Energy Physics, Chinese Academy of Sciences,
	Beijing 100049, China}
\affiliation{{College of Physics Sciences, University of Chinese Academy of Sciences, Beijing 100049, China}}

\begin{abstract}
Quantum electrodynamics is generally extended to a nonlocal QED by introducing the correlation functions. The gauge link is introduced to guarantee that the nonlocal QED is locally $U(1)$ gauge invariant. The corresponding Feynman rules as well as the proof of Ward-Takahashi identity are presented. As an example, the anomalous magnetic moments of leptons are studied in nonlocal QED. At one-loop level, besides the ordinary diagrams, there are many additional Feynman diagrams which are generated from the gauge link. It shows the nonlocal QED can provide a reasonable explanation for lepton $g-2$ anomalies.
\end{abstract}

\maketitle


\section{Introduction}
The standard model (SM) has been tested for a variety of experiments at great precision. In particular, the anomalous magnetic moment of lepton is one of the most precisely determined quantity in particle physics. The updated theoretical prediction of $a_\mu$ from standard model is $a_\mu^{\text{SM}}=116591810(43)\times10^{-11}$ \cite{Aoyama3}. The recent measurement of the muon anomalous magnetic moment by the E989 experiment at Fermilab shows 
\begin{equation}
\Delta a_\mu^{\text{FNAL}} = a_\mu^{\text{FNAL}} - a_\mu^{\text{SM}} = (230 \pm 69) \times 10^{-11},
\end{equation}
which is a $3.3 \sigma$ discrepancy from the SM prediction \cite{Abi}. Combined with the previous E821 experiment at BNL \cite{Bennett}, the result revealed a $4.2 \sigma$ deviation from the prediction of the SM \cite{Abi}
\begin{equation}
\Delta a_\mu = a_\mu^{\text{FNAL}+\text{BNL}} - a_\mu^{\text{SM}} = (251 \pm 59) \times 10^{-11}.
\end{equation}
For electron, the theoretical prediction of $a_e$ is $a_e^{\text{SM,B}}=1159652182.032(720)\times10^{-12}$ \cite{Aoyama}, where the superscript B means the fine structure constant $\alpha$ was measured at Berkeley with $^{137}{\text{Cs}}$ atoms \cite{Parker}. The most accurate measurement of $a_e$ has been carried out by the Harvard group and the discrepancy from SM was $2.4 \sigma$ \cite{Hanneke}
\begin{equation}\label{eq:e1}
\Delta a_e^{\text{B}}=a_e^{\text{exp}}-a_e^{\text{SM,B}}=(-87\pm36)\times10^{-14}.
\end{equation}
However, a new determination of the fine structure constant $\alpha$ \cite{Morel}, obtained from the measurement at Laboratoire Kastler Brossel (LKB) with $^{87}{\text {Rb}}$, improves the accuracy by a factor of $2.5$ compared to the previous best measurement at Berkeley \cite{Parker}. With this new $\alpha$, the SM prediction for the electron magnetic moment is $1.6 \sigma$ lower than the experimental data, i.e.,
\begin{equation}\label{eq:e2}
\Delta a_e^{\text{LKB}}=a_e^{\text{exp}}-a_e^{\text{SM,LKB}}=(48\pm30)\times10^{-14}.
\end{equation}
It is interesting that the two discrepancies of $\Delta a_e$ have similar size but opposite signs for still unidentified reasons \cite{Morel}. The small difference of $\alpha$ does not affect $\Delta a_\mu$ because it is much larger than $\Delta a_e$. As standard model predictions almost match perfectly all other experimental information, the deviation in one of the most precisely measured quantities in particle physics provides an enduring hint for new physics.

One can compare the anomalous magnetic moments between nucleons and leptons. Since the observation of the finite size of proton by Hofstadter in 1955 \cite{Hofstadter}, the electromagnetic form factors of proton and neutron have been widely studied for many decades. Theoretically, the anomalous magnetic moments of nucleon can be well described by effective field theory with the magnetic term explicitly included in the Lagrangian. However, the magnetic interaction between lepton and photon is incompatible with QED due to its non-renormalizability. Though there are some efforts to mitigate the discrepancy by correcting the theoretical calculation of the electron and muon anomalous magnetic moments in the standard model \cite{Borsanyi}, many works attempt to explain the muon $g-2$ result with new physics theories beyond the standard model. For example, the standard model was extended with an additional massive gauge boson $Z'$ which has new contributions to $(g-2)_\mu$ in Ref.~\cite{Ansta}. New particles directly couple to muon were introduced with the new physical contributions happening at one-loop level in Ref.~\cite{Balkin} while light millicharged particles were introduced with new contributions at two-loop level in Ref.~\cite{Bai} to explain the anomalies magnetic moment of muon. The authors of Ref.~\cite{Borah} provided a natural origin of $(g-2)_\mu$ anomaly in gauged $L_\mu$--$L_\tau$ model. Apart from the SM fermion content, the minimal version of this model has three heavy right handed neutrinos. In Refs.~\cite{Li,Wang}, the authors revisited constrained low energy supersymmetry (SUSY) models to match the $(g-2)_\mu$ result. There are also some other theories such as supergravity unified models \cite{Aboubrahim}, two Higgs doublet model \cite{Dey,Arcadi} and quark-lepton unification theory \cite{Perez,Ban}, etc explaining the $(g-2)_\mu$ result. Ref.~\cite{Athron} is a review of new physics explanations of $(g-2)_\mu$ anomaly with up to three new fields. 

Most of the above theoretical explanations focus on the $(g-2)_\mu$ anomaly and it is somewhat challenging to explaining both the muon and electron $g-2$ anomalies together in the same beyond-standard-model, because of their large magnitude difference and possible opposite signs. There have been some new physical models which attempt to explain the muon and electron $g-2$ anomalies simultaneously, such as supersymmetry model \cite{Endo,Badziak,JjCao}, new flavor model \cite{Calibbi}, gauged $U(1)_{e-\mu}$ extension \cite{Chen}, two Higgs doublet model \cite{Chun,Li2,Han,Botella}, type-III seesaw model \cite{seesaw}, 3-3-1 model \cite{seesaw2} and dark $Z_d$ model \cite{Cadeddu}, etc. The low-energy effective field theory and the standard model effective field theory were also applied to explain the lepton $g-2$ anomalies \cite{Aebischer}. In addition, in Ref.~\cite{Crivellin} $SU(2)_L$ doublet and singlet vector-like heavy leptons was introduced to couple to the SM leptons via Yukawa interaction to accommodate the anomalies in the electron and muon anomalous magnetic moments. The deviations of both $(g-2)_e$ and $(g-2)_\mu$ were explained by introducing one CP-even real scalar coupled to both electron and muon with different couplings in Ref.~\cite{Davoudiasl}. A complex singlet scalar was introduced in Ref.~\cite{Liu}, where the CP-odd scalar couples to electron only and the CP-even part couples to both muons and electrons. The authors of Ref.~\cite{Dutta} considered a general framework of the scalar singlet-doublet extension of the SM scalar sector and added three sterile neutrinos so that the anomalous magnetic moments of both muon and electron can be explained by the tree-level flavor violating couplings of the light scalar to the leptons. Ref.~\cite{Dorsner} investigated all possible ways to explain both $(g-2)_e$ and $(g-2)_\mu$ with either a single scalar leptoquark or a pair of scalar leptoquarks. Since the new introduced particles have nearly no visible effects to other physical observations, they are usually related to the candidates for dark matter \cite{Saez,Cox,Chak,Ghorbani,yag}.

One can see that theoretical solutions to the discrepancy of lepton magnetic moments were proposed without exception by introducing new particles, symmetries and interactions beyond standard model. We tried to explain the lepton $g-2$ anomalies from another way with nonlocal QED where no new particles were introduced \cite{g-2}. Nonlocal QED was inspired from nonlocal effective field theory (EFT) which reflects the non-point behavior of hadrons. The nonlocal EFT has been applied to study the nucleon electromagnetic form factors, strange form factors, parton distributions, etc \cite{h1,h2,h3,h4,h5,h6,h7,Review}. Though QED was proved to be the fundamental theory for electromagnetic interaction, the current precise measurements on lepton anomalous magnetic moments indicate leptons could also have ``structure". Nonlocal behavior could be general for all the interactions. In our previous work, the ``minimal" nonlocal extension of QED has been applied to explain the lepton $g-2$ anomalies \cite{g-2}. The unique advantage of this approach is that the Lagrangian has the same gauge symmetry and same interaction as QED, except it is nonlocal. The correlation functions in the nonlocal strength tension and lepton-photon interaction make it possible to explain the discrepancies of $\Delta a_\mu$ and $\Delta a_e$ simultaneously without introducing any new particles. 

In this work, we are keen on constructing a more general nonlocal QED theory based on our previous work \cite{g-2}. In Sec.~\ref{sec-2}, we will introduce the general extension of the local QED Lagrangian, where both the free and interaction parts are nonlocal. The Feynman rules for the vertices including the additional interaction generated from gauge link will be presented. In Sec.~\ref{sec-3}, we will prove that the Ward-Takahashi identity and charge conservation are satisfied for the nonlocal Lagrangian. All the self-energy and vertices diagrams at one-loop level will be included. The lepton anomalous magnetic moments will be studied with the nonlocal QED and numerical results will be discussed in Sec.~\ref{sec-4}. And finally Sec.~\ref{sec-5} is a short summary.

\section{Nonlocal QED and Feynman rules}\label{sec-2}
The local QED Lagrangian is written as
\begin{equation}
\mathcal{L}^{\text{local}}=\bar{\psi}(x)\lr{i\slashed{\partial}-m}\psi(x)-e\bar{\psi}(x)\slashed{A}(x)\psi(x)-\frac14F^{\mu\nu}(x)F_{\mu\nu}(x).
\end{equation}
Based on the same $U(1)$ symmetry, the local QED Lagrangian can be transformed into a nonlocal one using the method in \cite{v2,h1,h2,h3,h4,h5,h6,h7,v3,terning,g-2,terning2,Review}. The most general nonlocal Lagrangian can be written as 
\begin{align}
\mathcal{L}^{\text{nl}}&=\int d^4 a\bar{\psi}\lr{x+\frac{a}{2}}\bar{I}\lr{x,x+\frac{a}{2}}\lr{i\slashed{\partial}-m}\psi\lr{x-\frac{a}{2}}I\lr{x,x-\frac{a}{2}}F_1(a)\notag\\
&\quad-e\int d^4 a d^4 b\bar{\psi}\lr{x+\frac{a}{2}}\bar{I}\lr{x,x+\frac{a}{2}}\slashed{A}(x+b)\psi\lr{x-\frac{a}{2}}I\lr{x,x-\frac{a}{2}}F_1(a)F_2(a,b)\notag\\
&\quad-\frac14\int d^4 dF^{\mu\nu}(x)F_{\mu\nu}(x+d)F_4(d),
\label{eq-Lagrangian}
\end{align}
where the gauge link 
\begin{equation}\label{eq-link}
I(x,y)\equiv\text{exp}\lr{ie\int d^4 c\int_{x}^{y}d z^\mu A_\mu(z+c)F_3(a,c)}
\end{equation}
is introduced to guarantee the local gauge invariance. Compared with our previous work \cite{g-2}, here both the free Lagrangian and the interaction part are nonlocal in this general form of Eq.~\eqref{eq-Lagrangian}. The fermion fields $\psi$ and $\bar{\psi}$ are located at $x-\frac{a}{2}$ and $x+\frac{a}{2}$, respectively. The photon field $A_\mu$ is located at different coordinate $x+b$. The functions $F_1(a)$, $F_2(a,b)$, $F_3(a,c)$ and $F_4(d)$ are the correlation functions normalized as
\begin{equation}\label{eq-normalize}
\int d^4aF_1(a)=\int d^4bF_2(a,b)=\int d^4cF_3(a,c)=\int d^4dF_4(d)=1.
\end{equation}
Certainly, $\mathcal{L}^{\text{nl}}$ will turn back to $\mathcal{L}^{\text{local}}$ if $F_1(a)=\delta(a)$, $F_2(a,b)=\delta(b)$ and $F_4(d)=\delta(d)$. 
It is straightforward to prove the general nonlocal QED Lagrangian of Eq.~\eqref{eq-Lagrangian} is invariant under the following gauge transformation
\begin{equation}\label{eq-gaugetrans}
\psi(x)\to e^{i\alpha(x)}\psi(x),\quad A_\mu(x)\to A_\mu(x)-\frac{1}{ e}\partial_\mu\alpha'(x),
\end{equation}
where 
\begin{equation}\label{eq-alpha}
\alpha(x)=\int d^4 b\alpha'(x+b)F_2(a,b)=\int d^4 c\alpha'(x+c)F_3(a,c). 
\end{equation}

\begin{figure}[t]
\centering
\includegraphics{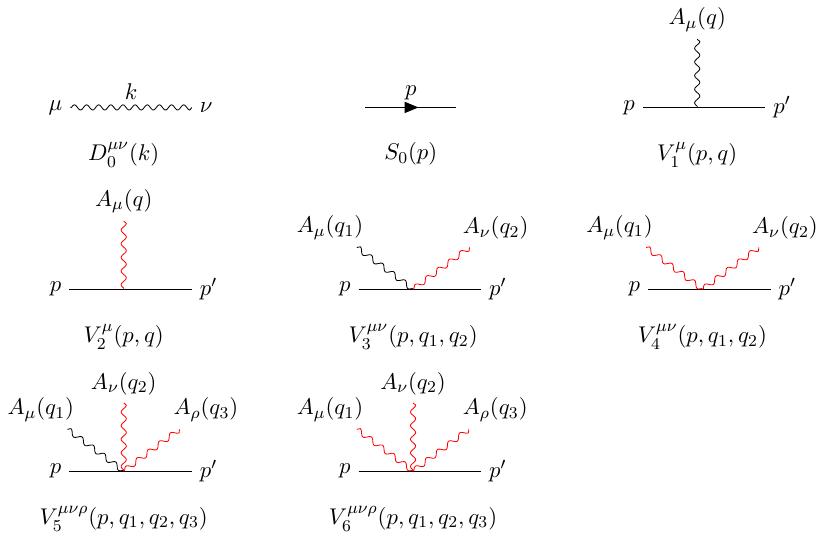}
\caption{Propagators and vertices in nonlocal QED appeared in one-loop diagrams. The black and red wavy lines are for photons from minimal substitution and gauge link, respectively. }\label{fig-v}
\end{figure}

With the nonlocal Lagrangian, one can derive the corresponding Feynman rules. The propagators and vertices are plotted in Fig.~\ref{fig-v}, where the black and red wavy lines in the vertices are for photons generated from the minimal substitution and gauge link, respectively. It is obvious that two or more photons can be generated from gauge link, while the minimal substitution can only generate one photon. Therefore in Fig.~\ref{fig-v}, the vertex could have one or more red photons, while it can only have one black photon at most. Since the free Lagrangian of fermion and photon fields in Eq.~\eqref{eq-Lagrangian} is nonlocal, their free propagators are modified as
\begin{gather}
S_0(p)=\frac{i}{\slashed{p}-m}\frac{1}{\tilde{F}_1(p)}
~~\text{and}~~ 
D_0^{\mu\nu}(k)=\frac{-i g^{\mu\nu}}{k^2}\frac{1}{\tilde{F}_4(k)},
\end{gather}
where $\tilde{F}_1(p)$ and $\tilde{F}_4(k)$ are Fourier transformations of the correlation functions $F_1(a)$ and $F_4(d)$, respectively. The normal fermion-photon interaction term in the nonlocal Lagrangian $\mathcal{L}^{\text{nl}}$ generated from the minimal substitution is $\int d^4 a d^4 b\bar{\psi}\lr{x+\frac{a}{2}}\slashed{A}(x+b)\psi\lr{x-\frac{a}{2}}F_1(a)F_2(a,b)$. 
The corresponding interaction vertex is expressed as
\begin{equation}\label{eq:v1}
V_1^\mu(p,q)=\gamma^\mu \int\frac{d^4 k}{(2\pi)^4}\tilde{F}_1(k)\tilde{F}_2\lr{\frac{p+p'}{2}-k,q}=\gamma^\mu\tilde{G}_{2}(p,q),
\end{equation}
where $\tilde{G}_{i}(p,q)$ is defined as
\begin{equation}
\tilde{G}_{i}(p,q)\equiv\tilde{G}_{i}(P,q)=\int\frac{d^4 k}{(2\pi)^4}\tilde{F}_1(k)\tilde{F}_i\lr{P-k,q}.
\end{equation}
The momentum $P$ is defined as $P\equiv\frac{p+p'}{2}$. $q$, $p$ and $p'$ are the momentum of photon, initial and final momentum of fermion, respectively.

Besides the normal interaction, the nonlocal Lagrangian $\mathcal{L}^{\text{nl}}$ has many additional interactions where the photons are generated from the gauge link \eqref{eq-link}. The method for deducing the Feynman rules of these vertices including gauge particles has been shown in Refs.~\cite{terning,terning2,v3,Review}. The related action for the interaction with one photon from the gauge link can be written as
\begin{equation}
S=-ie \int d^4ad^4cd^4x\bar{\psi}\left(x+\frac{a}{2}\right)G_3(a,c)\lr{i\slashed{\partial}_x-m}\psi\lr{x-\frac{a}{2}}I\lr{x-\frac a2, x+\frac a2},
\end{equation}
where $G_3(a,c)=F_{1}(a)F_3(a,c)$ and $I\lr{x-\frac a2, x+\frac a2}=\int_{x-\frac a2}^{x+\frac a2}d z^\mu A_\mu(z+c)$.
To get the Feynman rule for this vertex, we need to calculate $\int d^4 a d^4 c G_3(a,c)e^{iPa} I\lr{x-\frac a2, x+\frac a2}$.
Using the identity 
\begin{eqnarray}
\int d^4 a d^4 c G_3(a,c)  e^{iPa} I\lr{x-\frac a2, x+\frac a2} &=& \int d^4 a d^4 c d^4 k_1 d^4 k_2 \tilde{G}_3(k_1,k_2) e^{ik_1a}e^{ik_2c}e^{iPa} I\lr{x-\frac a2, x+\frac a2} \nonumber \\
&=& \int d^4 a d^4 c d^4 k_1 d^4 k_2 \left(\tilde{G}_3(-i\partial_a ,k_2) e^{ik_1a}\right)e^{ik_2c}e^{iPa} I\lr{x-\frac a2, x+\frac a2}
\end{eqnarray}
and partial integration, it is crucial to calculate $\tilde{G}_{3}(-i\partial_a,k_2)e^{iPa} I\lr{x-\frac a2, x+\frac a2}$.
Following the derivation of Refs.~\cite{terning,v3,Review}, one can show that
\begin{equation}
\tilde{G}_{3}(-i\partial_a,k_2)e^{iPa} I\lr{x-\frac a2, x+\frac a2} = e^{iPa} \tilde{G}_{3}(-i\mathcal{D}_a,k_2) I\lr{x-\frac a2, x+\frac a2},
\end{equation}
where $\mathcal{D}_a=\partial_a + iP_a$. Using the Taylor expansion and iteration method \cite{terning,terning2,v3,Review}, one can finally have
\begin{equation}
\tilde{G}_{3}(-i\partial_a,k_2) I\lr{x-\frac a2, x+\frac a2} = i\int d^4q \frac{P^\mu+q^\mu/2}{2P\cdot q +q^2}\left[\tilde{G}_3(p+q,q)-\tilde{G}_3(p,q)\right]\left[A(q)e^{iq(x+a/2)}+A(q)e^{iq(x-a/2)}\right].
\end{equation}
Therefore, the additional electromagnetic vertex with one photon from the gauge link is obtained as
\begin{equation}\label{eq:v2}
V_2^\mu(p,q)=(\slashed{p}-m)\frac{(q+2P)^\mu}{(q+P)^2-P^2}\left[\tilde{G}_{3}(q+p,q)-\tilde{G}_{3}(p,q)\right].
\end{equation}
The additional electromagnetic vertex with two photons (one from minimal substitution with momentum $q_1$ and the other from the gauge link with momentum $q_2$) can be obtained similarly as \cite{v3,terning,terning2}
\begin{align}\label{eq:v3}
V_3^{\mu\nu}(p,q_1,q_2)=i\gamma^\mu\frac{(q_2+2P)^\nu}{(q_2+P)^2-P^2}\left[\tilde{G}_{23}(q_2+p,q_1,q_2)-\tilde{G}_{23}(p,q_1,q_2)\right],
\end{align}
where $G_{ij}$ is defined as
\begin{equation}\label{eq-Gij}
\tilde{G}_{ij}(p,q_1,q_2)=\int\frac{d^4 k_1d^4 k_2}{(2\pi)^8}\tilde{F}_1(k_1)\tilde{F}_i\lr{k_2,q_1}\tilde{F}_j\lr{P-k_1-k_2,q_2}.
\end{equation}
The interaction vertex where two photons are both from the gauge link is expressed as
\begin{align}\label{eq:v4}
&V_4^{\mu\nu}(p,q_1,q_2)=i(\slashed{p}-m)\left\{2g^{\mu\nu}\frac{\tilde{G}_{33}(p+q_1+q_2,q_1,q_2)-\tilde{G}_{33}(p,q_1,q_2)}{(P+q_1+q_2)^2-P^2}\right.\notag\\
&-\frac{\tilde{G}_{33}(p+q_1+q_2,q_1,q_2)-\tilde{G}_{33}(p,q_1,q_2)}{(P+q_1+q_2)^2-P^2}\left[\frac{(2P+q_1)^\mu(2P+2q_1+q_2)^\nu}{(P+q_1+q_2)^2-(P+q_1)^2}+(\mu\leftrightarrow \nu, q_1\leftrightarrow q_2)\right]\notag\\
&\left.+\left[\frac{\left[\tilde{G}_{33}(p+q_1,q_1,q_2)-\tilde{G}_{33}(p,q_1,q_2)\right](2P+q_1)^\mu(2P+2q_1+q_2)^\nu}{\left((P+q_1)^2-p^2)((P+q_1+q_2)^2-(P+q_1)^2\right)}+(\mu\leftrightarrow \nu, q_1\leftrightarrow q_2)\right]\right\}.
\end{align}
In the above and following equations, when $q_i\leftrightarrow q_j$, only the first argument in the function $\tilde{G}$ makes such kind of change.
There are higher order interactions with more photons generated from the expansion of the gauge link. In this manuscript, we will study the lepton anomalous magnetic moments at one-loop level. The interactions up to three photons are needed. The interaction vertex with three photons where one is from minimal substitution and the other two are from the gauge link is obtained as
\begin{align}\label{eq:v5}
&V_5^{\mu\nu\rho}(p,q_1,q_2,q_3)=\gamma^\mu\left\{2g^{\nu\rho}\frac{\tilde{G}_{233}(p+q_2+q_3,q_1,q_2,q_3)-\tilde{G}_{233}(p,q_1,q_2,q_3)}{(P+q_2+q_3)^2-P^2}\right.\notag\\
&-\frac{\tilde{G}_{233}(p+q_2+q_3,q_1,q_2,q_3)-\tilde{G}_{233}(p,q_1,q_2,q_3)}{(P+q_2+q_3)^2-P^2}\left[\frac{(2P+q_2)^\nu(2P+2q_2+q_3)^\rho}{(P+q_2+q_3)^2-(P+q_2)^2}+(\nu\leftrightarrow \rho, q_2\leftrightarrow q_3)\right]\notag\\
&\left.+\left[\frac{\left[\tilde{G}_{233}(p+q_2,q_1,q_2,q_3)-\tilde{G}_{233}(p,q_1,q_2,q_3)\right](2P+q_2)^\nu(2P+2q_2+q_3)^\rho}{((P+q_2)^2-P^2)((P+q_2+q_3)^2-P^2)}+(\nu\leftrightarrow \rho, q_2\leftrightarrow q_3)\right]\right\},
\end{align}
where $\tilde{G}_{ijk}(p,q_1,q_2,q_3)$ is defined as
\begin{equation}
\tilde{G}_{ijk}(p,q_1,q_2,q_3)=\int\frac{d^4 k_1d^4 k_2d^4 k_3}{(2\pi)^{12}}\tilde{F}_1(k_1)\tilde{F}_i\lr{k_2,q_1}\tilde{F}_j\lr{k_3,q_2}\tilde{F}_k\lr{P-k_1-k_2-k_3,q_3}.
\end{equation}
The vertex for three photons which are all from the gauge link is more complicated. We can separate it into two terms as 
\begin{equation}\label{eq:v6}
V_6^{\mu\nu\rho}(p,q_1,q_2,q_3) = V_{6,a}^{\mu\nu\rho}(p,q_1,q_2,q_3) + V_{6,b}^{\mu\nu\rho}(p,q_1,q_2,q_3),
\end{equation}
where $V_{6,a}^{\mu\nu\rho}(p,q_1,q_2,q_3)$ is proportional to $g^{\mu\nu}$ expressed as
\begin{align}
&V_{6,a}^{\mu\nu\rho}(p,q_1,q_2,q_3)=2(\slashed{p}-m)g^{\mu\nu}\left\{-\frac{[\tilde{G}_{333}(p+q_1+q_2,q_1,q_2,q_3)-\tilde{G}_{333}(p,q_1,q_2,q_3)](2P+2q_1+2q_2+q_3)^\rho}{[(P+q_1+q_2)^2-P^2][(P+q_1+q_2+q_3)^2-(P+q_1+q_2)^2]}\right.\notag\\
&+\frac{\tilde{G}_{333}(p+q_1+q_2+q_3,q_1,q_2,q_3)-\tilde{G}_{333}(p,q_1,q_2,q_3)}{(P+q_1+q_2+q_3)^2-P^2}\left[\frac{(2P+2q_1+2q_2+q_3)^\rho}{(P+q_1+q_2+q_3)^2-(P+q_1+q_2)^2}\right.\notag\\
&+\frac{(2P+q_3)^\rho}{(P+q_1+q_2+q_3)^2-(P+q_3)^2}\bigg]
-\frac{[\tilde{G}_{333}(p+q_3,q_1,q_2,q_3)-\tilde{G}_{333}(p,q_1,q_2,q_3)](2P+q_3)^\rho}{[(P+q_3)^2-P^2][(P+q_1+q_2+q_3)^2-(P+q_3)^2]}\Bigg\}.
\end{align}
The other term $V_{6,b}^{\mu\nu\rho}(p,q_1,q_2,q_3)$ is expressed as
\begin{align}
&V_{6,b}^{\mu\nu\rho}(p,q_1,q_2,q_3)=(\slashed{p}-m)\Bigg\{\frac{\tilde{G}_{333}(p+q_1+q_2+q_3,q_1,q_2,q_3)-\tilde{G}_{333}(p,q_1,q_2,q_3)}{(P+q_1+q_2+q_3)^2-P^2}\notag\\
&\quad\times \bigg[\frac{(2P+2q_2+2q_3+q_1)^\mu(2P+2q_3+q_2)^\nu(2P+q_3)^\rho}{[(P+q_1+q_2+q_3)^2-(P+q_2+q_3)^2][(P+q_1+q_2+q_3)^2-(P+q_3)^2]}+(\mu\leftrightarrow \nu, q_1\leftrightarrow q_2)\bigg]\notag\\
&-\left[\frac{\tilde{G}_{333}(p+q_2+q_3,q_1,q_2,q_3)-\tilde{G}_{333}(p,q_1,q_2,q_3)}{(P+q_2+q_3)^2-P^2}\right. \notag\\
&\quad\times\left.\frac{(2P+2q_2+2q_3+q_1)^\mu(2P+2q_3+q_2)^\nu(2P+q_3)^\rho}{[(P+q_1+q_2+q_3)^2-(P+q_2+q_3)^2][(P+q_2+q_3)^2-(P+q_3)^2]}
+(\mu\leftrightarrow \nu, q_1\leftrightarrow q_2)\right]\notag\\
&-\frac{\tilde{G}_{333}(p+q_3,q_1,q_2,q_3)-\tilde{G}_{333}(p,q_1,q_2,q_3)}{(P+q_3)^2-P^2}\bigg[
\frac{(2P+2q_2+2q_3+q_1)^\mu(2P+2q_3+q_2)^\nu(2P+q_3)^\rho}{[(P+q_1+q_2+q_3)^2-(P+q_2+q_3)^2][(P+q_1+q_2+q_3)^2-(P+q_3)^2]}\notag\\
&\quad\left.-\frac{(2P+2q_2+2q_3+q_1)^\mu(2P+2q_3+q_2)^\nu(2P+q_3)^\rho}{[(P+q_1+q_2+q_3)^2-(P+q_2+q_3)^2][(P+q_2+q_3)^2-(P+q_3)^2]}+(\mu\leftrightarrow \nu, q_1\leftrightarrow q_2)\right]\notag\\
&+(\mu\to\nu,\nu\to\rho,\rho\to\mu,q_1\to q_2,q_2\to q_3,q_3\to q_1)+(\mu\to\rho,\nu\to\mu,\rho\to\nu,q_1\to q_3,q_2\to q_1,q_3\to q_2)\Bigg\}.
\end{align}
With the above Feynman rules, one can calculation the lepton magnetic form factors with the nonlocal Lagrangian.

\section{Charge conservation}\label{sec-3}
Before we start to calculate the magnetic moments, we first show the Ward-Takahashi identity and charge conservation can be obtained with the nonlocal Lagrangian. The nonlocal Lagrangian is invariant under the local $U(1)$ transformation with the following equation
\begin{equation}\label{eq-suppose1}
\int d^4xd^4a\bar{\psi}\left(x+\frac{a}{2}\right)\psi\left(x-\frac{a}{2}\right)F_1(a)\alpha(x)=\int d^4xd^4ad^4b\bar{\psi}\left(x+\frac{a}{2}\right)\psi\left(x-\frac{a}{2}\right)F_1(a)F_2(a,b)\alpha'(x+b).
\end{equation}
Eq.~(\ref{eq-alpha}) is obtained from the above requirement. For the fermions with momenta $k_1$ and $k_2$, one can get the following relationship from Eq.~(\ref{eq-suppose1}) as
\begin{equation}\label{eq-suppose2}
\tilde{F}_1(K)\tilde{\alpha}(k_1-k_2)=\int d^4k_3\tilde{F}_1(k_3)\tilde{F}_2(K-k_3,k_2-k_1)\tilde{\alpha}'(k_1-k_2),
\end{equation}
where $K\equiv\frac{k_1+k_2}{2}$. In Particular, when $k_1=k_2$, we have
\begin{equation}\label{eq-suppose3}
\tilde{F}_1(k_1)\tilde{\alpha}(0)=\int d^4k_3\tilde{F}_1(k_3)\tilde{F}_2(k_1-k_3,0)\tilde{\alpha}'(0).
\end{equation}
Meanwhile, the Fourier transformations of Eq.~\eqref{eq-normalize} and Eq.~\eqref{eq-alpha} are expressed as
\begin{equation}
\int d^4k\tilde{F}_2(k,0)e^{-i k\cdot a}=1
\end{equation}
and
\begin{equation}
\tilde{\alpha}(k)=\int d^4k'\tilde{F}_2(k',-k)\tilde{\alpha}'(k)e^{-i k'\cdot a},
\end{equation}
respectively. As a result, we have $\tilde{\alpha}(0)=\tilde{\alpha}'(0)$. Therefore, Eq.~(\ref{eq-suppose3}) can be rewritten as
\begin{equation}\label{eq-suppose4}
\tilde{F}_1(p)=\int d^4k\tilde{F}_1(k)\tilde{F}_2(p-k,0)\equiv G_2(p,q=0).
\end{equation}
In the same way, we also have 
\begin{equation}\label{eq-suppose5}
\tilde{F}_1(p)=\int d^4k\tilde{F}_1(k)\tilde{F}_3(p-k,0)\equiv G_3(p,q=0).
\end{equation}
With the definition of $G_{ij}$ in Eqs.~(\ref{eq-Gij}), one can get the relationship by using Eqs.~(\ref{eq-suppose4}) and (\ref{eq-suppose5}) as
\begin{align}
\tilde{G}_{ij}(p,q_1,q_2=0)&=\int\frac{d^4 k_1 d^4 k_2}{(2\pi)^8}\tilde{F}_1(k_1)\tilde{F}_i\lr{k_2,q_1}\tilde{F}_j\lr{p-k_1-k_2,0}
=\tilde{G}_{i}(p,q_1).\label{eq-Gij-Gi}
\end{align}
Similarly,
\begin{equation}\label{eq-Gijk-Gij}
\tilde{G}_{ijk}(p,q_1,q_2,q_3=0)=\tilde{G}_{ij}(p,q_1,q_2).
\end{equation}
Note that the above Eqs.~\eqref{eq-Gij-Gi} and \eqref{eq-Gijk-Gij} are valid for arbitrary one $q_i=0$. 

With the above equations, it is tedious but straightforward to obtain the following identities
\begin{figure}[t]
\centering
\includegraphics[scale=0.9]{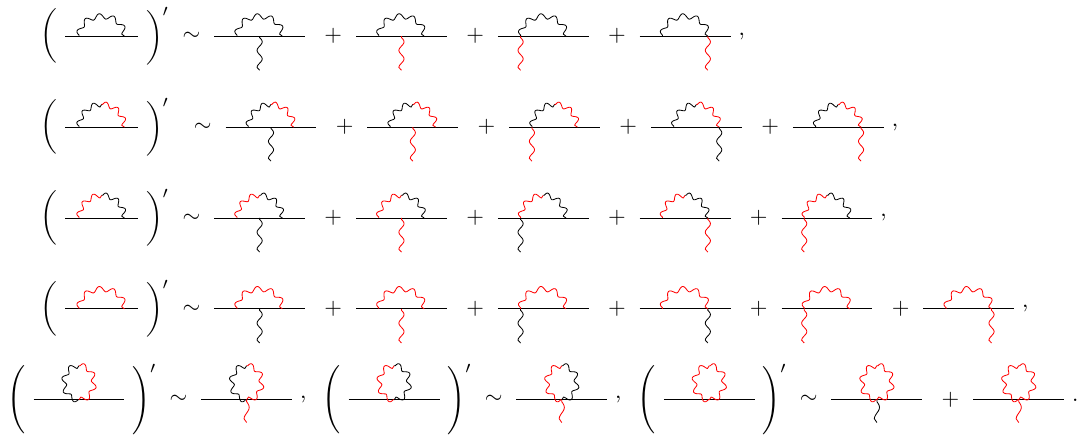}
\caption{Relations between lepton's self-energy diagrams and lepton-photon vertex diagrams  at one-loop level in nonlocal QED. The black and red wavy lines are for photons from minimal substitution and gauge link, respectively.}
\label{fig-ward}
\end{figure}
\begin{align}
\frac{ d S_0(p)}{ d p_\mu}&=i\lim_{q\to0}S_0(p)[V_1^\mu(p,q)+V_2^\mu(p,q)]S_0(p),\label{eq-d1}\\
\frac{\partial V_1^\mu(p,q_1)}{\partial p_\nu}&=i\lim_{q_2\to0}V_3^{\mu\nu}(p,q_1,q_2),\label{eq-d2}\\
\frac{\partial V_2^\mu(p,q_1)}{\partial p_\nu}&=i\lim_{q_2\to0}[V_3^{\nu\mu}(p,q_2,q_1)+V_4^{\mu\nu}(p,q_1,q_2)],\label{eq-d3}\\
\frac{\partial V_3^{\mu\nu}(p,q_1,q_2)}{\partial p_\rho}&=i\lim_{q_3\to0}V_5^{\mu\nu\rho}(p,q_1,q_2,q_3),\label{eq-d4}\\
\frac{\partial V_4^{\mu\nu}(p,q_1,q_2)}{\partial p_\rho}&=i\lim_{q_3\to0}[V_5^{\rho\mu\nu}(p,q_3,q_1,q_2)+V_6^{\mu\nu\rho}(p,q_1,q_2,q_3)]\label{eq-d5}.
\end{align}
Based on the above identities, the relationship between self-energy and vertex can be established. At one-loop level, there are 7 self-energy diagrams (Fig.~\ref{fig-self}) and 24 vertex diagrams (Fig.~\ref{fig-1loop}) plotted in Appendix \ref{app-B}. The total self-energy $\Sigma(p)$ and vertex $\Gamma^\mu(p,q)$ are expressed as
\begin{equation}
\Sigma(p) = \sum_{i=1}^7 \Sigma_i(p) ~~\text{and}~~ \Gamma^\mu(p,q) = V_1 + \sum_{i=1}^{24} \Gamma_i^\mu(p,q),
\end{equation}
where $\Sigma_i(p)$ and $\Gamma_i^\mu(p,q)$ are all written in Appendix \ref{app-B}. One can prove that 
\begin{align}
-\frac{ d \Sigma_1(p)}{ d p_\mu}&=\lim_{q\to0}[\Gamma^\mu_1(p,q)+\Gamma^\mu_2(p,q)+\Gamma^\mu_3(p,q)+\Gamma^\mu_5(p,q)],\\
-\frac{ d \Sigma_2(p)}{ d p_\mu}&=\lim_{q\to0}[\Gamma^\mu_6(p,q)+\Gamma^\mu_7(p,q)+\Gamma^\mu_8(p,q)+\Gamma^\mu_9(p,q)+\Gamma^\mu_{11}(p,q)],\\
-\frac{ d \Sigma_3(p)}{ d p_\mu}&=\lim_{q\to0}[\Gamma^\mu_{4}(p,q)+\Gamma^\mu_{12}(p,q)+\Gamma^\mu_{13}(p,q)+\Gamma^\mu_{14}(p,q)+\Gamma^\mu_{16}(p,q)],\\
-\frac{ d \Sigma_4(p)}{ d p_\mu}&=\lim_{q\to0}[\Gamma^\mu_{10}(p,q)+\Gamma^\mu_{15}(p,q)+\Gamma^\mu_{17}(p,q)+\Gamma^\mu_{18}(p,q)+\Gamma^\mu_{19}(p,q)+\Gamma^\mu_{20}(p,q)],\\
-\frac{ d \Sigma_5(p)}{ d p_\mu}&=\lim_{q\to0}\Gamma^\mu_{22}(p,q),\\
-\frac{ d \Sigma_6(p)}{ d p_\mu}&=\lim_{q\to0}\Gamma^\mu_{21}(p,q),\\
-\frac{ d \Sigma_7(p)}{ d p_\mu}&=\lim_{q\to0}[\Gamma^\mu_{23}(p,q)+\Gamma^\mu_{24}(p,q)].
\end{align}
These relations can be seen clearly from Fig.~\ref{fig-ward}. The derivative of the self-energy diagrams will result in the vertex diagrams. The derivative generates one external photon field which will be attached to the self-energy diagram at all possible places. For example, for the first rainbow self-energy diagram in Fig.~\ref{fig-ward}, a black photon from minimal substitution or a red photon from the gauge link will be attached to the internal lepton line, while only a red photon can be attached to the vertex. For the fourth rainbow diagram, both black and red photons can be attached to the vertex.

The dressed propagator $S(p)$ can be obtained as $S(p) = S_0(p) + S_0(p)\Sigma(p)S_0(p) + S_0(p)\Sigma(p)S_0(p)\Sigma(p)S_0(p) + \cdots$. With the above equations, it is straightforward to get
\begin{equation}\label{eq-q0loop}
\lim_{q\to0}S(p+q)[-i\Gamma^\mu(p+q,p)]S(p)=-\frac{dS(p)}{d p_\mu}.
\end{equation}
This indicates the Ward--Takahashi identity
\begin{equation}\label{eq-Ward}
\lim_{q\to0}\left[-iq_\mu\Gamma^\mu(p+q,p)\right]=\lim_{q\to0}\left[S^{-1}(p+q)-S^{-1}(p)\right].
\end{equation}
The dressed propagator can be also written as
\begin{equation}
S(p)=\frac{iZ_2}{\slashed{p}-m},
\end{equation}
where $Z_2$ is the wave function renormalization constant and $Z_2-1 = \frac{d\Sigma(p)}{d\slashed{p}}|_{\slashed{p}=m}$. Ward--Takahashi identity of Eq.~\eqref{eq-Ward} implies
\begin{equation}
Z_2\Gamma^\mu(p,p)=\gamma^\mu.
\end{equation}
The Dirac and Pauli form factors are defined as \cite{pes}
\begin{equation}
Z_2\Gamma^\mu(p+q,p)=\gamma^{\mu}F_1(q^2)+\frac{i\sigma^{\mu\nu}q_\nu}{2m}F_2(q^2).
\end{equation}
Hence $F_1(0)=1$. This is consistent with that the renormalized lepton charge is $1$. In the next section, we will use the nonloal QED to calculate the lepton magnetic moments to see whether the $g-2$ anomaly can be understood.

\section{Lepton anomalous magnetic moments}\label{sec-4}
By inserting the electromagnetic current into the lepton states with initial and final momenta $p$ and $p'$, one can get the Dirac and Pauli form factors of lepton. The corresponding Feynman diagrams are plotted in Fig.~\ref{fig-1loop} in Appendix \ref{app-B}. With the projection method, one can obtain
\begin{equation}
F_1=-\frac{A(4m^2-q^2)-6m^2B}{4(4m^2-q^2)^2},\quad F_2=m^2\frac{A(4m^2-q^2)-B(2m^2+q^2)}{q^2(4m^2-q^2)^2},
\end{equation}
where $A$ and $B$ are calculated from the following equations
\begin{equation}
A=\sum_{\text{spin}}\bar{u}(p')\Gamma_\mu u(p)\bar{u}(p)\gamma^\mu u(p')=\tr\left[\Gamma_{\mu}(\slashed{p}+m)\gamma^{\mu}(\slashed{p}'+m)\right]
\label{eq-A}
\end{equation}
and
\begin{equation}
B=\sum_{\text{spin}}\bar{u}(p')\Gamma_\mu u(p)\bar{u}(p)\frac{(p+p')^\mu}{m}u(p')=\tr\left[\Gamma_{\mu}(\slashed{p}+m)(\slashed{p}'+m)\frac{(p+p')^\mu}{m}\right].\label{eq-B}
\end{equation}
In this manuscript, we focus on the Pauli form factor $F_2$ which is related to the anomalous magnetic moment of lepton. Compared with the QED, in the nonlocal case, the one-loop vertices are much more complicated. In addition to the first diagram of Fig.~\ref{fig-1loop}, there are other 23 diagrams. For the first rainbow diagram, the vertex $\Gamma_1^\mu(p,q)$ is expressed as
\begin{equation}
\Gamma_1^\mu(p,q)=-e^2\int\frac{d^4k}{(2\pi)^4}V_1^\nu\left(p'-k,k\right)S_0(p-k+q)V_1^\mu(p-k,q)S_0(p-k)V_1^\rho\left(p,-k\right)D_{0\nu\rho}(k).
\end{equation}
The corresponding Pauli form factor of vertex $\Gamma_1^\mu(p,q)$, $F_{2,1}(q^2)$ is obtained as
\begin{align}
F_{2,1}(q^2)=&\frac{-8ie^2m^2}{q^2(4m^2-q^2)^2}\int\frac{d^4k}{(2\pi)^4}\left[\frac{(4m^2+2q^2)((k\cdot p)^2+(k\cdot p')^2)-8(m^2-q^2)(k\cdot p)(k\cdot p')}{((p'-k)^2-m^2)((p-k)^2-m^2)k^2}\right.\notag\\
&+\left.\frac{(q^4-4m^2q^2)(k\cdot p+k\cdot p'+k^2)}{((p'-k)^2-m^2)((p-k)^2-m^2)k^2}\right]\tilde{G}_{2}(p'-k,k)\tilde{G}_{2}(p-k,q)\tilde{G}_{2}(p,-k).
\end{align}
When momentum transfer $q^2=0$, one can get 
\begin{equation}
F_{2,1}(0)=ie^2 \int \frac{d^4k}{(2\pi)^4} \frac{(3k^2 + 2k \cdot p)m_l^2-3(k \cdot p)^2}{2k^2(k^2 - 2k \cdot p)^2m_l^2}\tilde{G}_{2}(p-k,k)\tilde{G}_{2}(p-k,0)\tilde{G}_{2}(p,-k).
\end{equation}
For all the other diagrams, the expressions of $\Gamma_i^\mu(p,q)$ are written in Appendix \ref{app-B}.

In the numerical calculation, we treat photon as a point particle and will not modify the photon propagator. The correlation function $F_4(a)$ in the free photon Lagrangian is chosen to be $\delta(a)$. For the lepton-photon interaction, the function $F_i(a,b)~(i=2,3)$ is assumed to be factorized as $f(a)F_i(b)$ in order to simplify numerical calculation. According to Eq.~(\ref{eq-normalize}), $f(a)$ should be an $a$-independent constant 1 and $\int d^4 b F_i(b)=1$. As a result, the Fourier transformations of the correlators can be expressed as
\begin{equation}
\tilde{G}_i(p,q)=\tilde{F}_1(P)\tilde{F}_i(q).
\end{equation}
For the vertex with two and three photons, they can be factorized similarly as
\begin{align}
\tilde{G}_{ij}(p,q_1,q_2)&=\tilde{F}_1(P)\tilde{F}_i(q_1)\tilde{F}_j(q_2),\\
\tilde{G}_{ijk}(p,q_1,q_2,q_3)&=\tilde{F}_1(P)\tilde{F}_i(q_1)\tilde{F}_j(q_2)\tilde{F}_k(q_3),
\end{align}
where $i$, $j$ and $k$ can be 2 or 3. The subscript 2 is for photon from minimal substitution and 3 for photon from the gauge link.
For the correlators in the interaction vertex, they are chosen to be
\begin{equation}
\tilde{F}_2(k)=\tilde{F}_3(k)=\frac{\Lambda_2^2}{\Lambda_2^2-k^2}.
\end{equation}
These correlators have been proposed in our earlier work on the nonlocal effective field theory and ``minimal" version of nonlocal QED \cite{Review,g-2}. The difference here is that the free Lagrangian of lepton is also nonlocal which results in a modified lepton propagator. For $\tilde{F}_1(p)$ in the free lepton propagator, it is chosen to be
\begin{equation}\label{eq-regulator1}
\tilde{F}_1(p)=\frac{\Lambda_1^2-p^2}{\Lambda_1^2}.
\end{equation}
Since the correlator is in the denominator of the propagator, the modified propagator makes the loop integral more convergent. 

In the above correlators, $\Lambda_1$ and $\Lambda_2$ are free parameters. With the parameters, the difference between nonlocal QED and standard model $\Delta a_l^{\text{nl}}$ can be obtained. We determine these cutoff parameters $\Lambda$s with the experimental $\Delta a_e$ and $\Delta a_\mu$. 
We should address that nonlocal QED itself can not pre-determine the form of the correlation functions and the parameters $\Lambda_1$ and $\Lambda_2$. They reflect the properties of the particles and should be determined by experiments. Practically, one can try and compare different correlators. The parameters $\Lambda$s will be determined to reproduce the experimental data. In this case, we determine $\Lambda_{2e}$ and $\Lambda_{2\mu}$ for a given $\Lambda_1$ with the experimental $\Delta a_e$ and $\Delta a_\mu$ for the chosen correlators. 

\begin{figure}[h]
    \centering
    \includegraphics{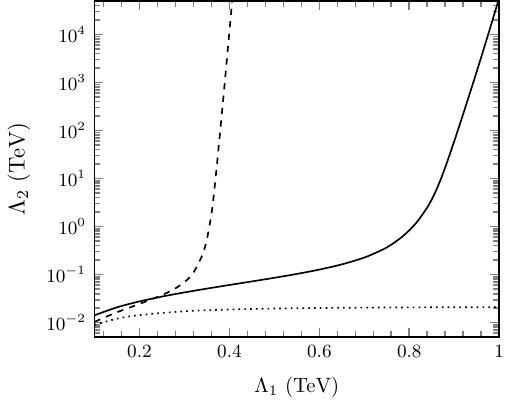}
    \caption{The cutoff parameter $\Lambda_2$ versus $\Lambda_1$. The solid, dashed and dotted lines are for muon discrepancy $\Delta a_\mu$, electron discrepancies $\Delta a_e^{\text{LKB}}$ and $\Delta a_e^{\text{B}}$, respectively.}
    \label{fig:Lambda_2-Lambda_1-1}
\end{figure}

In Fig.~\ref{fig:Lambda_2-Lambda_1-1}, the obtained $\Lambda_2$ versus $\Lambda_1$ are plotted. The solid, dashed and dotted lines are for muon discrepancy $\Delta a_\mu$, electron discrepancies $\Delta a_e^{\text{LKB}}$ and $\Delta a_e^{\text{B}}$, respectively. 
At each vertex, the correlator $\tilde{F}_1(k)$ makes the integral more divergent, while $\tilde{F}_2(k)$ and $\tilde{F}_3(k)$ make the integral more convergent. For a given $\Lambda_1$, one can always find a corresponding $\Lambda_2$ to get the experimental discrepancies. For muon, when $\Lambda_1$ is small, say less than 0.8 TeV, $\Lambda_2$ increases smoothly with the increasing $\Lambda_1$. Meanwhile $\Lambda_2$ increases rapidly when $\Lambda_1$ is larger than about 0.8 TeV. For example, for $\Lambda_1=0.8$ TeV, 0.9 TeV and 1.0 TeV, the corresponding $\Lambda_2$ are $0.84$ TeV, $60.23$ TeV and $5.82\times 10^4$ TeV, respectively. In order to make the nonlocal effect negligible in other electromagnetic processes, we prefer large value of $\Lambda$s (at least hundreds of GeV). For electron case, the results for the two discrepancies $\Delta a_e^{\text{LKB}}$ and $\Delta a_e^{\text{B}}$ are quite different. For the negative $\Delta a_e^{\text{B}}$, the obtained $\Lambda_2^{\text{B}}$ is not sensitive to $\Lambda_1$. At very broad range of $\Lambda_1$, the value of $\Lambda_2^{\text{B}}$ is around $10-20$ GeV, which is too small to be reasonable. The processes, for example, the cross section $e^+ e^- \rightarrow \mu^+ \mu^-$ will be different from the experiments due to small cutoff parameter $\Lambda_2$. However, for the new result of $\Delta a_e^{\text{LKB}}$, when $\Lambda_1$ is about 0.35 TeV, the obtained $\Lambda_2^{\text{LKB}}$ is around 0.55 TeV. The corresponding $\Lambda_2^{\text{LKB}}$ increases very quickly with the increasing $\Lambda_1$ when $\Lambda_1$ is larger than 0.35 TeV. Since for any $\Lambda_1$, the anomalous magnetic moment of electron is infinite when $\Lambda_2$ goes to infinity, one can always get a corresponding $\Lambda_2^{\text{LKB}}$ to obtain $\Delta a_e^{\text{LKB}}$ for any large $\Lambda_1$.

Therefore, the muon $g-2$ discrepancy can be well explained with large $\Lambda$s, which will not cause contradiction with the other experimental measurements. But for the electron case, $\Delta a_e^{\text{B}}$ can not be reasonably reproduced in nonlocal QED. As we have pointed, the correlators in the vertex $\tilde{F}_1(k)$ and $\tilde{F}_i(k)$ $i=2,3$ have opposite effects. With proper choice of $\Lambda$s we can still get negative $\Delta a_e^{\text{nl}}$ with large $\Lambda$s, though its absolute value is much smaller than $\Delta a_e^{\text{B}}$. For example, the calculated $\Delta a_e^{\text{nl}}$ is $-4.04\times10^{-16}$ with $\Lambda_2=1.0$ TeV and infinite $\Lambda_1$. Certainly, for a given $\Lambda_1$, if $\Lambda_2$ is between $\Lambda_2^{\text{B}}$ and $\Lambda_2^{\text{LKB}}$, the calculated $\Delta a_e^{\text{nl}}$ is between $\Delta a_e^{\text{B}}$ and $\Delta a_e^{\text{LKB}}$. For $\Lambda_1=1.0$ TeV, if we assume $\Lambda_2$ for electron is the same as for muon, the discrepancy $\Delta a_e^{\text{nl}}$ is $8.19\times10^{-14}$.

\begin{figure}[hbt]
\centering
\includegraphics{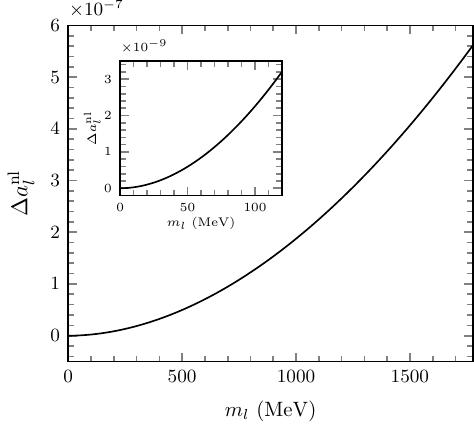}
\caption{The calculated discrepancy of lepton anomalous magnetic moment $\Delta a_l^{\text{nl}}$ versus lepton mass $m_l$. $\Lambda_1$ is fixed to be 1 TeV and $\Lambda_2$ is fixed to be $\Lambda_{2\mu}$ obtained by the experimental $\Delta a_\mu$ with $\Lambda_1=1$ TeV. The small figure at the corner is for the result at small lepton mass.}
\label{fig:a_l-m}
\end{figure}

To see clearly the lepton-mass dependence of the calculated discrepancy $\Delta a_l^{\text{nl}}$, in Fig.~\ref{fig:a_l-m}, we plot the $\Delta a_l^{\text{nl}}$ versus lepton mass $m_l$ with $\Lambda_1=1$ TeV. A small figure is plotted at the corner to show the result at small lepton mass. Because $\Lambda_{2}$ for electron is not well determined due to two different measurements, for any mass $m_l$, $\Lambda_2$ is chosen to obtain the experimental $\Delta a_\mu$ with $\Lambda_1=1$ TeV. Therefore, when $m_l=m_\mu$, $\Delta a_l^{\text{nl}}$ equals to the experimental discrepancy $\Delta a_\mu$. The calculated $\Delta a_l^{\text{nl}}$ is bigger than $\Delta a_e^{\text{B}}$ but smaller than $\Delta a_e^{\text{LKB}}$ when $m_l=m_e$. The discrepancy $\Delta a_l^{\text{nl}}$ increases with the increasing lepton mass and it is $5.62\times 10^{-7}$ when $m_l$ is at the mass of $\tau$ lepton. With the nonlocal QED, both the muon and electron $g-2$ anomalies can be reasonably explained. Different from other theoretical methods, we did not introduce any new symmetries and new particles. The lepton mass dependence of $\Delta a_l^{\text{nl}}$ is naturally reproduced. It also leads to a large positive discrepancy $\Delta a_\tau^{\text{nl}}$ which can be tested by precise experiments in the future.

\section{Summary}\label{sec-5}

In summary, we proposed a general form of nonlocal QED. With the inclusion of gauge link, the nonlocal Lagrangian is locally $U(1)$ gauge invariant. The Ward-Takahashi identity and charge conservation are also satisfied. The nonlocal Lagrangian leads to modified propagators and vertices. Besides the normal interaction from the minimal substitution, there are additional interactions generated from the gauge link. With the nonlocal QED which is one kind of extension of the standard model, we studied the anomalous magnetic moments of leptons. Both the experimental discrepancies $\Delta a_\mu$ and $\Delta a_e^{\text{LKB}}$ can be well understood with proper choice of $\Lambda_{1}$ and $\Lambda_2$ in the correlators. Since the cutoff parameter $\Lambda$s should be large enough, the discrepancy $\Delta a_e^{\text{B}}$ can not be reproduced, though we can still get negative $\Delta a_e^{\text{nl}}$ with smaller magnitude. If the cutoff parameters are assumed to be the same for electron and muon, the calculated $\Delta a_e^{\text{nl}}$ is between the two experimental values $\Delta a_e^{\text{B}}$ and $\Delta a_e^{\text{LKB}}$. In addition, nonlocal QED naturally gives the lepton mass dependence of the discrepancy $\Delta a_l^{\text{nl}}$. Different from almost all the other solutions of $g-2$ anomalies, our nonlocal QED did not introduce any new symmetries and particles.

\section*{Acknowledgments}

This work is supported by the NSFC under Grant No. 11975241.
\appendix

\section{Self energy and vertices at one-loop level}\label{app-B}
\begin{figure}[tbph]
\centering
\includegraphics{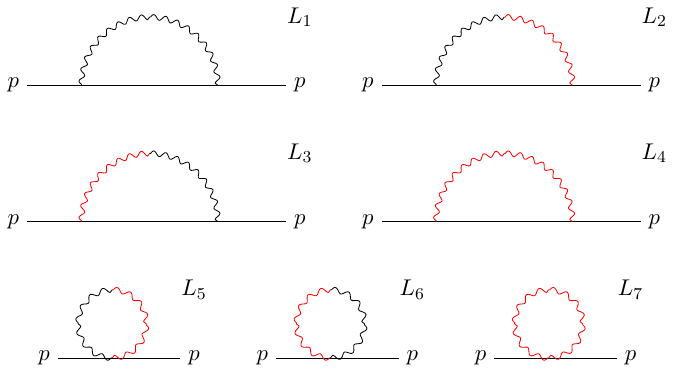}
\caption{Feynman diagrams of lepton self energy at one-loop level in nonlocal QED. The black and red wavy lines are for photons from minimal substitution and gauge link, respectively.}\label{fig-self}
\end{figure}

For the nonlocal QED Lagrangian there are 7 one-loop diagrams for lepton's self energy, as shown in Fig. \ref{fig-self}. The first diagram is similar as the local QED case where the lepton-photon interaction are obtained from the minimal substitution. The expression for the first diagram is written as 
\begin{equation}
-i\Sigma_1(p)=-e^2\int\frac{d^4k}{(2\pi)^4}V_1^\nu\left(p-k,k\right)S_0(p-k)V_1^\mu\left(p,-k\right)D_{0\mu\nu}(k),
\end{equation}
where $S_0$ and $D_{0\mu\nu}$ are propagators of lepton and photon. $V_1^\mu$ is the interacting vertex of Eq.~(\ref{eq:v1}). The diagrams of $L_2$ and $L_3$ include one photon from minimal substitution (black wavy line) and another photon from gauge link (red wavy line). The self-energy for these two diagrams are expressed as
\begin{equation}
-i\Sigma_2(p)=-e^2\int\frac{d^4k}{(2\pi)^4}V_2^\nu\left(p-k,k\right)S_0(p-k)V_1^\mu\left(p,-k\right)D_{0\mu\nu}(k)
\end{equation}
and 
\begin{equation}
-i\Sigma_3(p)=-e^2\int\frac{d^4k}{(2\pi)^4}V_1^\nu\left(p-k,k\right)S_0(p-k)V_2^\mu\left(p,-k\right)D_{0\mu\nu}(k),
\end{equation}
where $V_2^\mu$ is the interacting vertex of Eq.~(\ref{eq:v2}). The self-energy of the rainbow diagram $L_4$ is expressed as
\begin{eqnarray}
-i\Sigma_4(p)&=&-e^2\int\frac{d^4k}{(2\pi)^4}V_2^\nu\left(p-k,k\right)S_0(p-k)V_2^\mu\left(p,-k\right)D_{0\mu\nu}(k),
\end{eqnarray}
where the photons in both vertices are from the gauge link. The last three are bubble diagrams, where two photons (either from minimum substitution or from gauge link) can be emitted from one vertex. The self-energy of the bubble diagram $L_5$ and $L_6$ are expressed as
\begin{eqnarray}
-i\Sigma_5(p)&=&-e^2\int\frac{d^4k}{(2\pi)^4}V_3^{\mu\nu}(p,-k,k)D_{0\mu\nu}
\end{eqnarray}
and
\begin{eqnarray}
-i\Sigma_6(p)&=&-e^2\int\frac{d^4k}{(2\pi)^4}V_3^{\nu\mu}(p,k,-k)D_{0\mu\nu},
\end{eqnarray}
where $V_3^{\mu\nu}$ is the vertex of Eq.~(\ref{eq:v3}). The expression of the self-energy for the last bubble diagram $L_7$ is written as
\begin{eqnarray}
-i\Sigma_7(p)&=&-e^2\int\frac{d^4k}{(2\pi)^4}V_4^{\mu\nu}(p,k,-k)D_{0\mu\nu},
\end{eqnarray}
where $V_4^{\mu\nu}$ is the vertex of Eq.~(\ref{eq:v4}).

\begin{figure}[btph]
\centering
\includegraphics[scale=0.9]{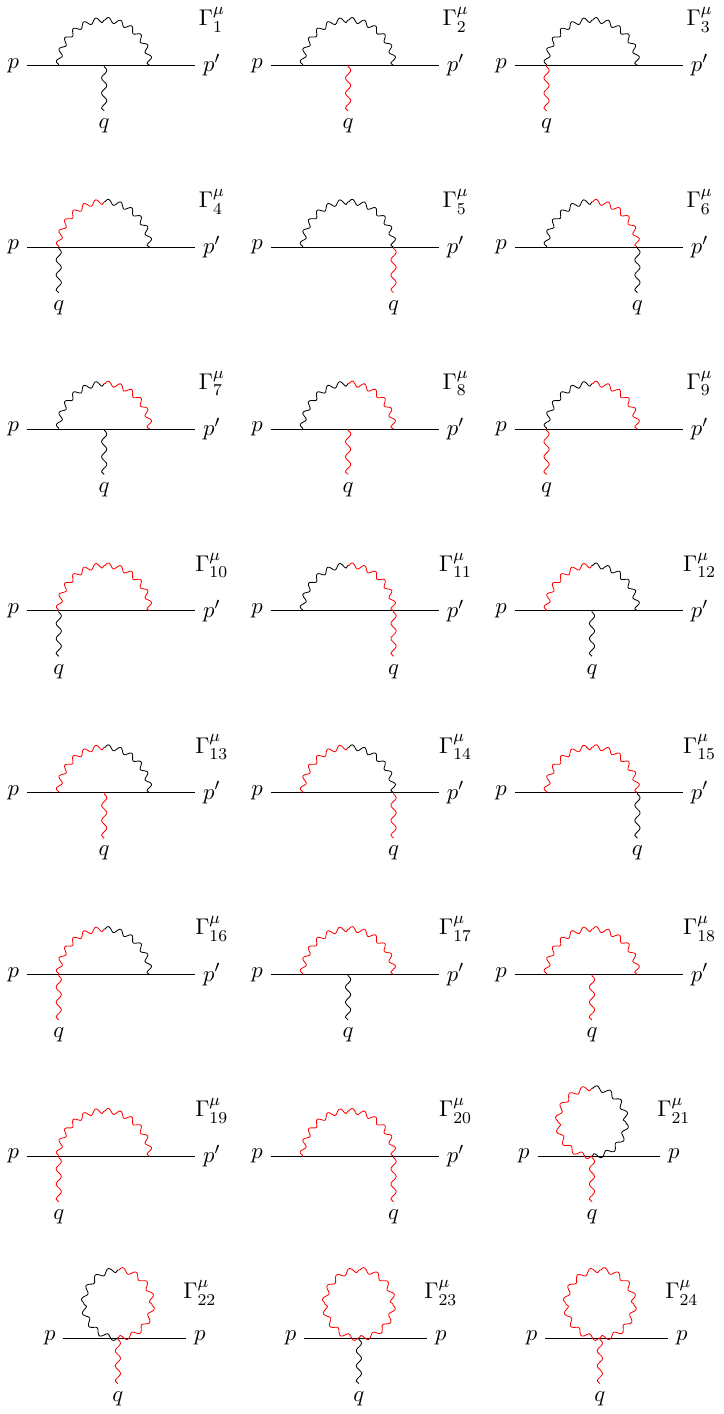}
\caption{Feynman diagrams for the vertices at one-loop level in nonlocal QED. The black and red wavy lines are for photons from minimal substitution and gauge link, respectively.}\label{fig-1loop}
\end{figure}

All the lepton-photon vertices $\Gamma_i^\mu(p,q)$ at one-loop level are plotted in Fig.~\ref{fig-1loop}. The first diagram is similar as the local QED and the expression of $\Gamma_1^\mu(p,q)$ is expressed as
\begin{eqnarray}
\Gamma_1^\mu(p,q)&=&-e^2\int\frac{d^4k}{(2\pi)^4}V_1^\nu\left(p'-k,k\right)S_0(p-k+q)V_1^\mu(p-k,q)S_0(p-k)V_1^\rho\left(p,-k\right)D_{0\nu\rho}(k).
\end{eqnarray}
The other 23 diagrams in Fig.~\ref{fig-1loop} contain additional vertices of nonlocal QED. In each diagram, both black photon (from minimal substitution) and red photon (from gauge link) can interact with lepton. At each interacting vertex, two or more photons can be emitted with only one black photon at most. The expressions of $\Gamma_i^\mu(p,q)$ for these 23 diagrams are written as
\begin{align*}
\Gamma_2^\mu(p,q)&=-e^2\int\frac{d^4k}{(2\pi)^4}V_1^\nu\left(p'-k,k\right)S_0(p-k+q)V_2^\mu(p-k,q)S_0(p-k)V_1^\rho\left(p,-k\right)D_{0\nu\rho}(k),\\
\Gamma_3^\mu(p,q)&=-e^2\int\frac{d^4k}{(2\pi)^4}V_1^\nu\left(p'-k,k\right)S_0(p-k+q)D_{0\nu\rho}(k)V_3^{\mu\rho}\left(p,-k,q\right),\\
\Gamma_4^\mu(p,q)&=-e^2\int\frac{d^4k}{(2\pi)^4}V_1^\nu\left(p'-k,k\right)S_0(p-k+q)D_{0\nu\rho}(k)V_3^{\mu\rho}\left(p,q,-k\right),\\
\Gamma_5^\mu(p,q)&=-e^2\int\frac{d^4k}{(2\pi)^4}V_3^{\mu\nu}\left(p-k,k,q\right)S_0(p-k)D_{0\nu\rho}(k)V_1^{\rho}\left(p,-k\right),\\
\Gamma_6^\mu(p,q)&=-e^2\int\frac{d^4k}{(2\pi)^4}V_3^{\mu\nu}\left(p-k,q,k\right)S_0(p-k)D_{0\nu\rho}(k)V_1^{\rho}\left(p,-k\right),\\
\Gamma_7^\mu(p,q)&=-e^2\int\frac{d^4k}{(2\pi)^4}V_2^\nu\left(p'-k,k\right)S_0(p-k+q)V_1^\mu(p-k,q)S_0(p-k)V_1^\rho\left(p,-k\right)D_{0\nu\rho}(k),\\
\Gamma_8^\mu(p,q)&=-e^2\int\frac{d^4k}{(2\pi)^4}V_2^\nu\left(p'-k,k\right)S_0(p-k+q)V_2^\mu(p-k,q)S_0(p-k)V_1^\rho\left(p,-k\right)D_{0\nu\rho}(k),\\
\Gamma_9^\mu(p,q)&=-e^2\int\frac{d^4k}{(2\pi)^4}V_2^\nu\left(p'-k,k\right)S_0(p-k+q)D_{0\nu\rho}(k)V_3^{\mu\rho}\left(p,-k,q\right),\\
\Gamma_{10}^\mu(p,q)&=-e^2\int\frac{d^4k}{(2\pi)^4}V_2^\nu\left(p'-k,k\right)S_0(p-k+q)D_{0\nu\rho}(k)V_3^{\mu\rho}\left(p,q,-k\right),\\
\Gamma_{11}^\mu(p,q)&=-e^2\int\frac{d^4k}{(2\pi)^4}V_4^{\mu\nu}\left(p-k,q,k\right)S_0(p-k)D_{0\nu\rho}(k)V_1^{\rho}\left(p,-k\right),\\
\Gamma_{12}^\mu(p,q)&=-e^2\int\frac{d^4k}{(2\pi)^4}V_1^\nu\left(p'-k,k\right)S_0(p-k+q)V_1^\mu(p-k,q)S_0(p-k)V_3^\rho\left(p,-k\right)D_{0\nu\rho}(k),\\
\Gamma_{13}^\mu(p,q)&=-e^2\int\frac{d^4k}{(2\pi)^4}V_1^\nu\left(p'-k,k\right)S_0(p-k+q)V_2^\mu(p-k,q)S_0(p-k)V_2^\rho\left(p,-k\right)D_{0\nu\rho}(k),\\
\Gamma_{14}^\mu(p,q)&=-e^2\int\frac{d^4k}{(2\pi)^4}V_3^{\mu\nu}\left(p-k,k,q\right)S_0(p-k)D_{0\nu\rho}(k)V_2^{\rho}\left(p,-k\right),\\
\Gamma_{15}^\mu(p,q)&=-e^2\int\frac{d^4k}{(2\pi)^4}V_3^{\mu\nu}\left(p-k,q,k\right)S_0(p-k)D_{0\nu\rho}(k)V_2^{\rho}\left(p,-k\right),\\
\Gamma_{16}^\mu(p,q)&=-e^2\int\frac{d^4k}{(2\pi)^4}V_1^\nu\left(p'-k,k\right)S_0(p-k+q)D_{0\nu\rho}(k)V_4^{\mu\rho}\left(p,q,-k\right),\\
\Gamma_{17}^\mu(p,q)&=-e^2\int\frac{d^4k}{(2\pi)^4}V_2^\nu\left(p'-k,k\right)S_0(p-k+q)V_2^\mu(p-k,q)S_0(p-k)V_2^\rho\left(p,-k\right)D_{0\nu\rho}(k),\\
\Gamma_{18}^\mu(p,q)&=-e^2\int\frac{d^4k}{(2\pi)^4}V_2^\nu\left(p'-k,k\right)S_0(p-k+q)V_2^\mu(p-k,q)S_0(p-k)V_2^\rho\left(p,-k\right)D_{0\nu\rho}(k),\\
\Gamma_{19}^\mu(p,q)&=-e^2\int\frac{d^4k}{(2\pi)^4}V_2^\nu\left(p'-k,k\right)S_0(p-k+q)D_{0\nu\rho}(k)V_4^{\mu\rho}\left(p,q,-k\right),\\
\Gamma_{20}^\mu(p,q)&=-e^2\int\frac{d^4k}{(2\pi)^4}V_4^{\mu\nu}\left(p-k,q,k\right)S_0(p-k)D_{0\nu\rho}(k)V_2^{\rho}\left(p,-k\right),\\
\Gamma_{21}^\mu(p,q)&=-e^2\int\frac{d^4k}{(2\pi)^4}V_5^{\nu\rho\mu}(p,k,-k,q)D_{0\nu\rho},\\
\Gamma_{22}^\mu(p,q)&=-e^2\int\frac{d^4k}{(2\pi)^4}V_5^{\rho\nu\mu}(p,-k,k,q)D_{0\nu\rho},\\
\Gamma_{23}^\mu(p,q)&=-e^2\int\frac{d^4k}{(2\pi)^4}V_5^{\mu\nu\rho}(p,q,k,-k)D_{0\nu\rho},\\
\Gamma_{24}^\mu(p,q)&=-e^2\int\frac{d^4k}{(2\pi)^4}V_6^{\mu\nu\rho}(p,k,-k,q)D_{0\nu\rho},\\
\end{align*}
where the interacting vertices $V_5^{\mu\nu\rho}$ and $V_6^{\mu\nu\rho}$ are written in Eqs.~(\ref{eq:v5}) and (\ref{eq:v6}). With the above expressions for the self-energy and vertices, one can prove the Ward-Takahashi identity as shown in the text.

\end{document}